\documentclass[runningheads,a4paper]{llncs}

\usepackage{amssymb}
\usepackage{graphicx}
\graphicspath{{./images/}}

\usepackage{multirow} 
\usepackage[lofdepth,lotdepth]{subfig}

\usepackage{url}
\urldef{\mailsa}\path|{bgliwa, kozlak, azygmunt, cetnar}agh.edu.pl|
\newcommand{\keywords}[1]{\par\addvspace\baselineskip
\noindent\keywordname\enspace\ignorespaces#1}

\begin{document}

\mainmatter  


\title{Models of~social groups in~blogosphere based~on information about comment addressees and~sentiments}

\titlerunning{Models of social groups in blogosphere \ldots}

%
%
\author{
Bogdan Gliwa \and Jaros{\l}aw Ko{\'z}lak \and Anna Zygmunt \and Krzysztof Cetnarowicz 
}
\authorrunning{}

\institute{
AGH University of Science and  Technology\\
Al. Mickiewicza 30, 30-059 Krak\'ow, Poland\\
\mailsa
}

%
%

\toctitle{}
\tocauthor{}
\maketitle

\begin{abstract}
\emph{
This work concerns the analysis of number, sizes and other characteristics of groups identified in the blogosphere using a set of models identifying social relations. These models differ regarding identification of social relations, influenced by methods of classifying the addressee of the comments (they are either the post author or the author of a comment on which this comment is directly addressing) and by a sentiment calculated for comments considering the statistics of words present and connotation. The state of a selected blog portal was analyzed in sequential, partly overlapping time intervals. Groups in each interval were identified using a version of the CPM algorithm, on the basis of them, stable groups, existing for at least a minimal assumed duration of time, were identified.
}

\end{abstract}

\section{Introduction}

An important problem in the analysis of social media is to identify the real relations between users in the best possible way, which allows us to identify groups that best reflect reality considering majority of existing significant interactions between entities and their emotional (sentimental) characteristics.

Nowadays, blogs play a significant role in the exchange of information on different subjects and the forming of opinions. A very important element of blogs is the possibility of adding comments, which facilitate discussions. 
Comments may be written in relation to posts or other comments and  may have a different content and emotional attitude.
Blogosphere is very dynamic, thus the relationships between bloggers are very dynamic and temporal: the lifetime of posts is very short.

In the research on blogosphere, different interactions between users are used for constructing models for analysis. 
This paper concerns the analysis of number, sizes and other characteristics of groups identified in the blogosphere using a set of models identifying social relations. 
These models differ regarding the method of classifying the addressee of the comments (they are either the post author or the author of a comment on which this comment is directly addressing) and a sentiment calculated for comments considering the statistics of words present and connotation.

Taking into consideration the sentiment while analysing groups allows us to identify groups built by interactions having different degrees of positive, neutral or negative sentiment.  Qualification of differences between such groups may be important not only for sociological research, but also for identification of kind of influence and its consequences, applied for example to choice of  advantageous marketing politics or identification of influential users who spread verbal violence and hatred.

\section{Research domain overview}

\subsection{Models of blogosphere}

The research concerning the analysis of blogosphere,  
produced constructions of different models of parts of blogosphere. 
One can observe that the character of these models is strictly dependent on the kinds of analysis for which they are created, e.g.  identification of key users and groups. 

For such applications, it is possible to distinguish universal models, which embrace both the representation of the character of given nodes and the links between them, the models focusing on the classification of nodes without considering the strength of the links between given pairs of nodes and models focusing mostly on neighborhoods of nodes and not taking the characteristic features of individual nodes into consideration. 

In \cite{AgarwalLiu:09} several graph structures related to blogs are distinguished: a blog network (formed by linked blogs), post network (formed by linked posts) and blogger network (formed by linked bloggers).
The authors consider different methods of identification of links between nodes: (i) hyperlinks to other blogs existing on the blogs,
 (ii) every pair of nodes, whose distance is smaller than a given constant $\epsilon$ are connected by links, (iii) number of $k$ nodes nearest to a given node is connected to it, (iv) all blogs are connected by edges with weights expressing similarities of given blogs.

Another important factor of the models is the dynamics of existing links and their weights in time.
 In  \cite{Ning:2010}, focused on the analysis of the evolving blog groups, 
the similarity relations between blogs were expressed, which  led to considering them as members of the same group.
 In \cite{Chi:2007} the authors proposed a method (community factorization) for representation of structures and temporal dynamics of blog groups.
In \cite{Akritidis:2009} a model for the identification of influential bloggers is presented, which took into consideration the time of interactions and when the given post ceased to be influential, causing new interactions to represent  links between blogs.

\subsection{Groups in social networks}

There are many definitions of groups (communities, clusters), mainly according to the area in which they were created. So it is difficult to find in literature an unequivocal definition of a group, acceptable to everybody \cite{Tang2010}. A group can be treated as a dense subset of vertices in a network, which are loosely connected with vertices outside the group. In practice, in complex social networks, groups are not isolated and individuals can be, in a given time, members of many groups.  
Many methods of finding groups (overlapping or not) have been proposed. In \cite{Fortunato2010} there are detailed descriptions of the most popular methods and algorithms. Every group can be described by several parameters, e.g. density (ratio of the number of links within the group to the maximum possible number of links), stability (the ratio of the number of people, present in both group to the number of all group members), cohesion (ratio of the average strength of links between the members to the average strength of their links with people outside the group).

Due to the nature of the blogosphere (the user may be a member of various discussion groups), the most useful are the algorithms finding overlapping groups. The most prominent representative of this group of methods is CPM algorithm \cite{palla2005,palla2008} where groups are defined
as sub-graphs consisting of a set of connected k-cliques. With the increase of parameter k the smaller and more disintegrated groups arise \cite{palla2005} and there is a suggestion that values of k = 3,...,6 seem to be the most appropriate.  

\subsection{Sentiment analysis}

Emotions are an integral component of statements in social media, especially on blogs or forums. Different groups of users can discuss the same topics in a~~completely different atmosphere, supporting each other or disagreeing. For each such statement, we can assign a value expressing an emotional attitude: positive, negative, neutral, objective or bipolar \cite{Tromp2011}.  

A large increase in interest in problems of analysis of sentiment can be seen around 2001. Some reasons for such interest in this research area are shown in \cite{Pang2008}: the development of advanced methods of analysis of natural language, which were already mature enough that it can be successfully applied in practice, more and easier availability of test data that were suitable for such analyzes (mostly available on the WWW) and the increasing demand for intelligent applications.

The term ``sentiment analysis'' (also used later interchangeably with ``opinion mining'')  was initially pertained to ``automatic analysis of evaluative text and tracking of the predictive judgments'' and was closely associated with analyzing market sentiment. Later, 
the term was rather treated as classifying reviews according to their polarity: either positive or negative. Nowadays the term refers to ``computational treatment of
opinion, sentiment, and subjectivity in text'' \cite{Pang2008}.
Sentiment analysis is closely related to natural language processing. 
Analysis of sentiment generally consists of several steps (\cite{Tromp2011}): part-of-speech tagging (division into language tokens), subjectivity detection (determining the statement as subjective or objective) and polarity detection (for subjective statements evaluate their polarity). 
There are different techniques and statistical methodologies to evaluate sentiment. 

The main difficulty in assessing the sentiment is that it is context-sensitive.
Currently, the increasingly popular use of sentiment analysis is the analysis of political blogs \cite{Gryc2010}, and more recently Twitter \cite{Shamma2010} due to the high amounts of opinions, sentiments and emotions. 

\subsection{Sentiment analysis in domain of social networks and group identification}

The general idea of finding groups in a social network (e.g. blogosphere) is to identify a set of vertices, communicating to each other more frequently than with vertices outside the group, regardless of the expressed emotional potential. 
Simply counting the number of comments and the weight of edges connecting two users does not distinguish situations when a user writes a comment in support of the ideas expressed by another in a post and when he disagrees with the writer of the post he/she is commenting.

In \cite{Xu2011} authors focus on group detection based on links and sentiment -- they were finding non-overlapping clusters that share similar sentiment. The researchers claim that this is the first work on sentiment group detection.
 In this work, they propose two methods of finding such communities. The first method assumes that sentiment can be either positive or negative. In the second method, the range of sentiment is divided into intervals and group users into groups according to the specific differences in the ranges of values of sentiment.

The problem of sentiment based clustering was used directly for the analysis of the blogosphere in \cite{Nguyen2010}. The authors proposed an algorithm called hyper-community detection and they used two methods: content-based hyper-community detection and sentiment-based hyper-community detection. In the first, they extracted topics from blog content, while the second method used sentiment information (from mood tags or emotion words used in posts).

In paper \cite{Bermingham2009}, the authors use sentiment analysis with social network approach in the context of radicalisation, searching terrorists in some specific groups from  {the Youtube portal \footnote{www.youtube.com}}. They tried to find out whether a chosen group was populated by radicals who could convince others to their beliefs and whether males or females are more radical. Sentiment analysis was used to define the level of radicalization of their comments containing some chosen keywords and social network analysis -- to extract key members in the group and to compare some network characteristics between a male and a female group.
               In article \cite{Kraus2008} authors tried to predict the success in the Oscar Awards based on analysis of communication on {IMDb portal\footnote{the Internet Movie Database -- www.imdb.com}}. They used sentiment analysis as a tool to define positivity of the user's opinions about movies -- authors searched for positive keywords that were extracted based on their betweenness centrality. The researches took advantage of social network analysis by weighting user posts according to the importance, expressed by betweenness centrality, of users that wrote them and treating most influential users as people who can possibly create trends.

\section{Dynamic models of social system}

Our model of social system, which first version was presented in \cite{KozlakZ11}, is adapted to the analysis of the characteristics of groups, their formation, dynamic, reasons and predicted character of future evolution. The state of the system is analyzed in subsequent time intervals called time slots. For each such interval the interactions taking place between entities are analyzed, and groups identified. It is assumed, that the groups may overlap.

For the identification of the groups the Clique Percolation Method \cite{palla2005,palla2008} in the version for a directed graph with weights is used. Then, among such identified groups the stable groups are discovered, using  
SGCI (Stable Group Changes Identification) algorithm \cite{ZygmuntBKK11,ZygmuntBKK12,GliwaSZBKK12}.
The concept of stable groups was introduced  
due the dynamic character of blogosphere, where groups may change very rapidly, and for our analysis of the evolution of blogosphere the most interesting are groups which last for a longer time.
The condition that a group is considered as a stable group is to identify in the next time slots groups with similar sets of members, evaluated using the Jaccard measure modified by us (expressed as a ratio of size of intersection of the pair of considered groups to the size of one of the groups from them - the larger value of such a ratio is considered as the modified Jaccard measure). The group is stable if it has such similar groups at least during the minimum assumed number of time slots.

The model is described in two parts -- the first (described in section \ref{FunctionSocietyModel}) concerns the fundamental elements of the model -- entities and interactions among them (more details in section \ref{InteractionsBetweenBloggers}), and the second (section \ref{OrganisationModel}) -- the organization with social system (section \ref{BuildingOfSocialRelations}) and groups.

\subsection{Fundamental model of social system}
\label{FunctionSocietyModel}

Dynamic model of social system $Soc(t)$, describes its state in the time slot $t$: 

\begin{equation}
Soc(t) = (N(t), X(t), \zeta, I(t), Org(t))
\label{eq:society}
\end{equation}
where:
\begin{description} 
	\item $N(t)$ -- set of entities building a social system,
   \item $X(t)$ -- vectors of values of measures calculated for the entities from the set $N$,
                   $X_{N_i}(t)$ represents a vector of measures of the entity
                  $N_i$ for the time slot $t$,
	\item $\zeta$ -- function, which assigns values of a vector of measures 
                to entities $N$, 
	\item I(t) -- set of interactions, consists of all the interactions between entities, together with the times they took place, their type, sets of involved entities and their roles in the interaction, the content and/or sentiment of the exchanged information, 
	\item $Org(t)$ -- organization of the social system, described in section \ref{OrganisationModel}.
\end{description}

\subsection{Interactions between bloggers}
\label{InteractionsBetweenBloggers}

Applying the model to the analyzed blogosphere domain and the analyzed problem of group identification, we can distinguish the following kinds of interactions between entities: commenting on posts, commenting on a comment, static links in blogs or posts to another blog/post, logins/nicks of bloggers mentioned in the content of post or comment.
The identification of some of these mentioned interaction types burdened by the varying level of uncertainty, whether the assignment was correct or not. 
In our work we are focusing on the interactions caused by commenting on posts of other users or by commenting on previously written comments to posts.
These interactions have varying characters which make them useful while analyzing the dynamics of groups and for a significant part of them it is possible to correctly identify who is being addressed.

The representation of the individual interaction, assumed by us, is as follows: 
\begin{equation}
i_l  = (N_i, N_j, N_p, t_z, k, s) 
\label{eq:interaction}
\end{equation}
where: 
  $N_i$ -- interaction initiator (writer of post or comment), 
  $N_j$ --  the addressee of the comment (sometimes not specified), 
 $N_p$ -- author of post to which the comment/interaction is written, 
 $t_z$ --  given time slot,
  $k$ --  type, which may be post, comments to post, comments to comment, 
 $s$ --  sentiment value, expressed in the bounded interval [-1, 1].

\subsection{Organization of social system}
\label{OrganisationModel}

The organization of social system $Org$ is expressed using the following elements:
\begin{equation}
Org(t) = (R(t), \psi , GT(t), \gamma, G(t), \xi, XG(t), \zeta^g)
\label{eq:organisation}
\end{equation}
\begin{description}
	\item $R(t)$ -- social relation, shaped as the results of interactions taking place,
\item $\psi$ -- function which builds social relations $R$  between a pair of entities, on the basis of interaction taking place between them,  
\begin{equation}
R(a,b,t_z) = \psi(I(a, b, t_z))  
\label{eq:socialrelation}
\end{equation}
Equation (\ref{eq:socialrelation}) shows social relation between users $a$ and $b$ in the time slot $t_z$, $\psi$ returns a strength of the relation expressed as a positive real number.   

\item	$GT(t)$ -- set of identified temporary groups, 

\item $\gamma$ -- a function which assigns entities to fugitive groups, 
 $\gamma : N \times R \rightarrow GT \times \{0, 1\}$
The used method of the classifications of nodes to groups is as follow: for each time slot, the fugitive groups are identified on the basis of the version of the CPM algorithm, calculated for a directed graph with weights.

\item	$G(t)$ --  set of identified stable groups,
Groups are considered as stable, when their life span equals at least $ltmin$ (which is set in the tests as equal to 3). 

\item $\xi$ -- function which identifies stable groups among fugitive ones, $\xi: GT \rightarrow G$, 

\item $XG(t)$ -- vectors of values of measures calculated for the groups by $\zeta^g$ , $XG_{Gr_i}(t)$ represents a vector assigned to a group $Gr_i$ which may be temporary (element of $G$) or stable (element of $GT$),
\item	$\zeta^g$ -- a function which calculates values of defined vectors  of measures for temporary or stable groups and assigns it to 
$XG(t)$.

\end{description}

\subsection{Building of social relations}
\label{BuildingOfSocialRelations}

In our model of social relations two main factors are considered: the frequency of interactions between nodes and the sentiment of interactions. 
The sentiment of the interaction may be classified into one of three groups: positive interaction, negative interaction or neutral (indifferent) interaction, on the basis of content analysis and strength of positive or negative connotation of words appearing in the comment. 

In this work the following versions of the $\psi$ function are distinguished:
\begin{itemize}
\item $\psi_{pn}$ -- considers all comments as addressed to the author of post, does not take sentiment of comments into consideration,

\item $\psi_{cn}$ --
scores comments which have a defined addressee of the comments as addressed to this addressee and not to the post author, if it is not possible to identify the addressee, the comment is scored as addressed to the post author, sentiment is not taken into consideration,

\item $\psi_{cs}$ -- scores comments which have a defined addressee of the comments as addressed to this addressee and not to the post author, if it is not possible to identify the addressee, the comment is scored as addressed to the post author, sentiment is taken here into consideration, and 
either relations caused by each kind of the sentiment (positive, negative, neutral) are considered separately or average values of the sentiment for every existing links are calculated, making this link to appear only in that adequate kind of sentiment model. The following subversion can be distinguished:
\begin{itemize}

\item $\psi_{cs,p}$, $\psi_{cs,n}$, $\psi_{cs,i}$  (sentiment counting models) -- in the given models, only interactions with positive ($cs,p$), negative (cs,n) or neutral (cs,i) sentiment are considered, for every pair of users interactions with each sentiment are scored separately,
\item $\psi_{cs,p+i}$ -- similar to previous ones, interactions with positive or neutral sentiment are taken into consideration together, the interactions with negative sentiment are omitted,

\item $\psi_{cs,p}^a$, $\psi_{cs,n}^a$, $\psi_{cs,i}^a$  (sentiment mean models) -- the average value of sentiment for a given ordered pair of users  is taken into consideration,  the directed relation between two users may be assigned only to one of these (which means positive, negative and neutral) models,

\item $\psi_{cs,p+i}^a$ -- similar to previous ones, but considers links with both positive or neutral average sentiment.
\end{itemize}
\end{itemize}

\section{Application of models to group identification and analysis}

\subsection{Description of experiments}

{\bf Data set.}
The analyzed data set contains data from the portal www.salon24.pl which consists of blogs (mainly political, but also have subjects from different areas). The data set consists of 26 722 users (11 084 of them have their own blog), 285 532 posts and 4 173 457 comments within the period 1.01.2008 - 31.03.2012.  The analyzed period was divided into time slots, each lasting 30 days. The neighboring slots overlap each other by 50\% of their length and in the examined period there are 104 times lots.

The large graph from all time slots consists of 26 053 nodes and 663 098 edges. Nodes in this graph are the users - both the owners of blogs and people only commenting on other posts. The number of nodes in the graph is lower than the overall number of active authors (26 722) in the given period, because some posts did not have any comments. Thus their authors cannot appear in this graph, unless they had commented on others or had any of their posts commented on.

{\bf Data set preparation.} We decided to remove edges with weights below 2 
to eliminate some noise and to reduce calculation time.
After removing such edges, the number of nodes was equal to 
15 578 (59.8\% of initial number of nodes) and the number of edges to 311 718 (47\% of the initial number of edges). When we are considering the number of connections as the number of edges multiplied by their weights, then the removed edges constitute 8.42\% of such connections.

To extract groups from networks we used CPMd version (for directed graphs) of CPM from {CFinder\footnote{www.cfinder.org}} tool, for different k in ranges 3 to 5.

{\bf Sentiment calculation.}
The sentiment for posts and comments was calculated using a tool developed at  {the Luminis Research company\footnote{www.luminis-research.com}}. Their method is based on searching words from analyzed text in a dictionary and counting sentiment for found ones. The dictionary is manually built and contains about 37~000 words (including about 4000 positive and negative words together -- the others are the neutral ones). Each word in the dictionary has a weight in the range $<-1;1>$ -~ negative values determine negative sentiment, positive -- positive one and neutral words have a weight equal to zero (intensity of positive or negative sentiment depends on assigned value - the closer value to 1 or -1, the greater the intensity of the sentiment is). Then the sentiment values for found words in the dictionary are summed and using the sum value, the number of positive, negative and neutral words in analyzed text the final sentiment value is calculated (based on heuristic equation with mentioned values). The final value describing the overall sentiment is between -1 and 1, but thresholds for negative, neutral and positive sentiment need adjusting. This can be done by analyzing some texts (part of texts earlier marked by algorithm) by human, manually assigning sentiment values (positive/negative/neutral) for them, next comparing these values with algorithm ones and finally setting appropriate thresholds.

In order to adjust thresholds for sentiment values, we analyzed about 150 random texts and based on this analysis we set the following thresholds: negative ($< 0$), neutral ($0-0.3$), positive: ($> 0.3$).

\subsection{Comparison of post and comments models}

During the analysis of the groups emerging in the blogosphere it is very important to identify, at first, the real characters of the interactions taking place, especially who is sending and receiving them.

In the case of comments, although they are assigned to a given post, in reality they often refer directly to an earlier entry commenting in this post.
In the blog portal salon24 we analyzed, the identification of the receiver of the comments is not that evident as they are only assigned to the post and the commenter can only refer to the name of the bloggers whose comment they are  commenting on. But, this is not done in an automatic way, the blogger is only able to do it by appropriately writing the subject of their comment (by writing ``@bloggername'' there). It is not always common practice, and if not specified, the writer of any post is considered as a receiver of that comment. 

\begin{figure}[ht]
\centering
\includegraphics[scale=0.27]{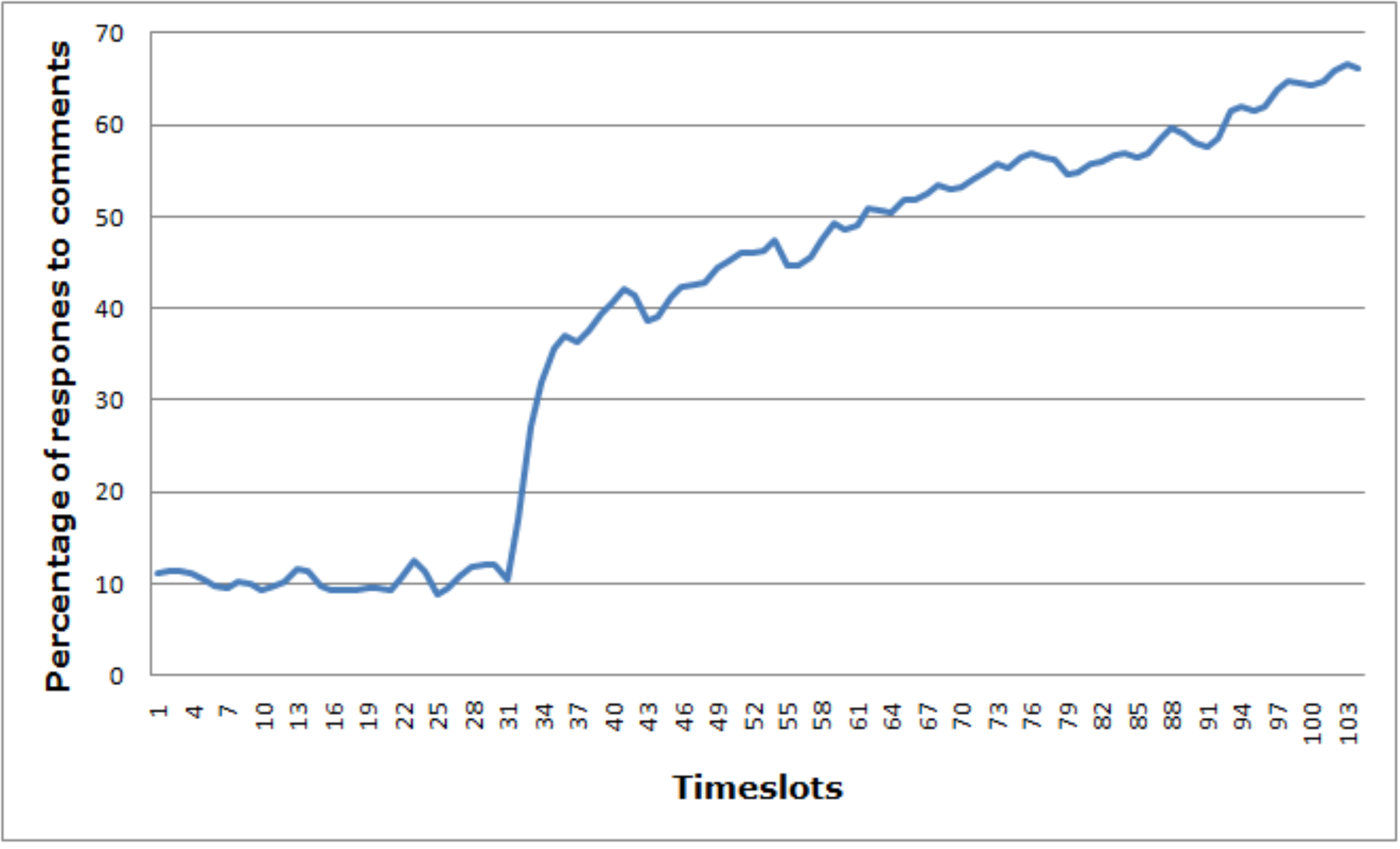}
\caption{Percentage of responses of type comment-comment to all responses.}
\label{fig:commentsRespones}
\end{figure}

For all 4~173~457 comments we identified 1~953~571 as comments that are responses to other comments (about 50~\%).
In fig. \ref{fig:commentsRespones} a noticeable increase in the percentage of comments having the receiver specified in such a way in time may be seen, so in the majority of cases it is possible to correctly consider that information in the model, what increases the accuracy of the represented interactions between bloggers and the subsequent emerging social relations.

\begin{figure}[t]
\centering
\subfloat[Number of groups in timeslots.]{
\includegraphics[scale=0.28]{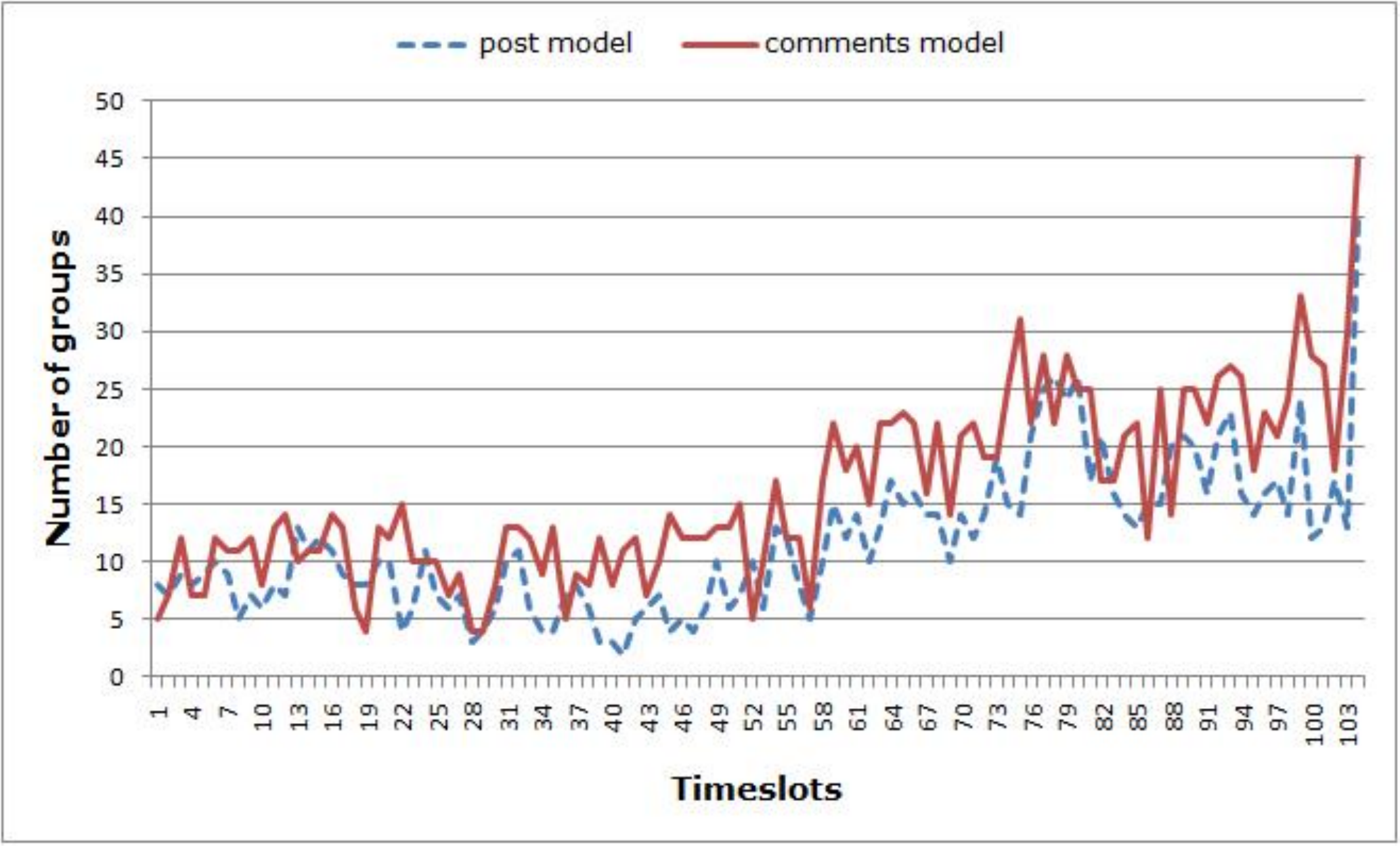}
\label{fig:grInSlots}
}
\subfloat[Users not belonging to any stable group.]{
\includegraphics[scale=0.28]{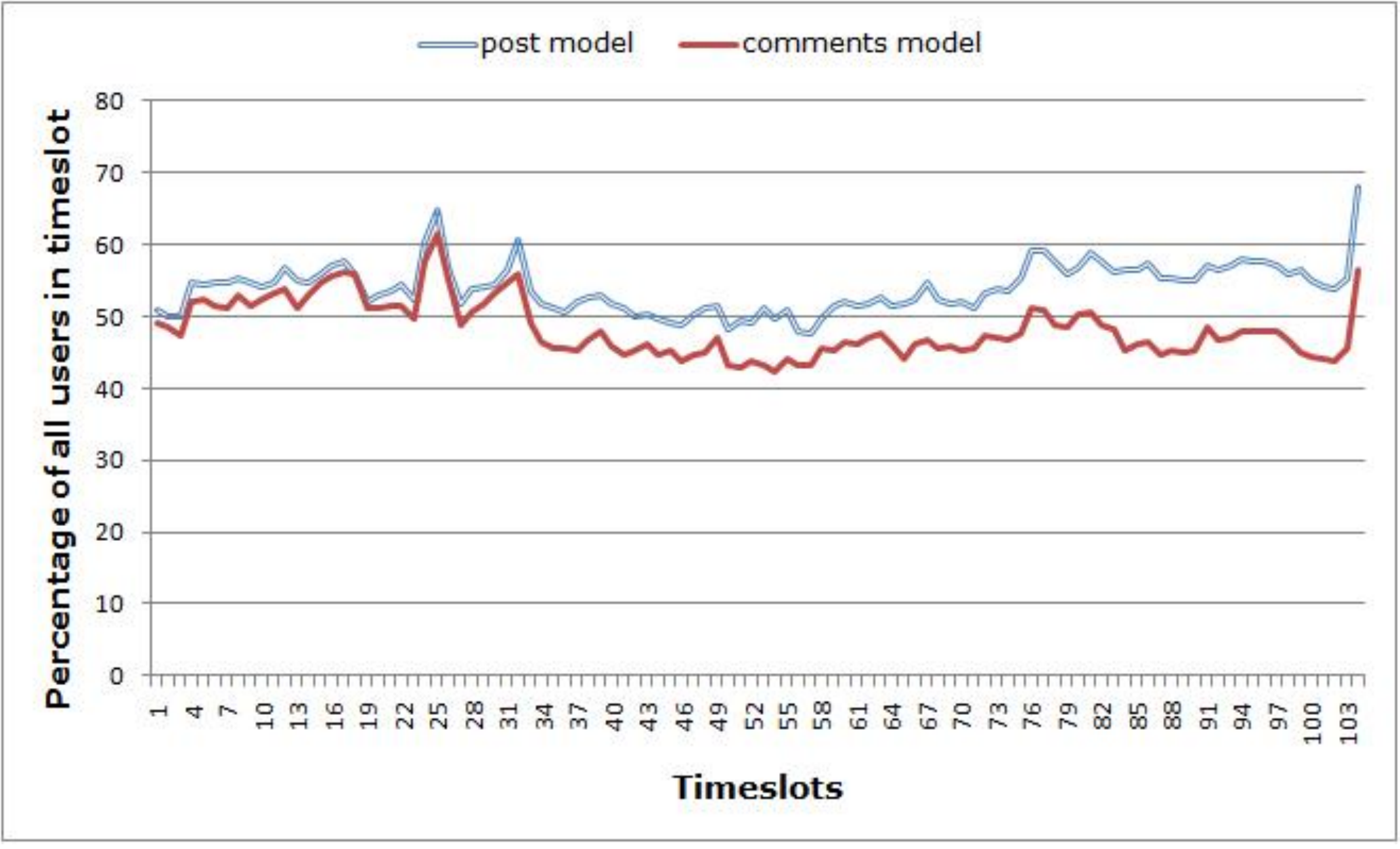}
\label{fig:notInGr}
}
\caption{Comparison between post and comments models for k=3.}
\end{figure}

Such assumptions are confirmed by the fact that in the new model (comments model $\psi_{cn}$) more groups were identified (see fig. \ref{fig:grInSlots}) than in old one (post model $\psi_{pn}$ ), a smaller part of user are not assigned to any groups (see fig. \ref{fig:notInGr}).

\begin{figure}[t]
\centering
\subfloat[1 group]{
\includegraphics[scale=0.28]{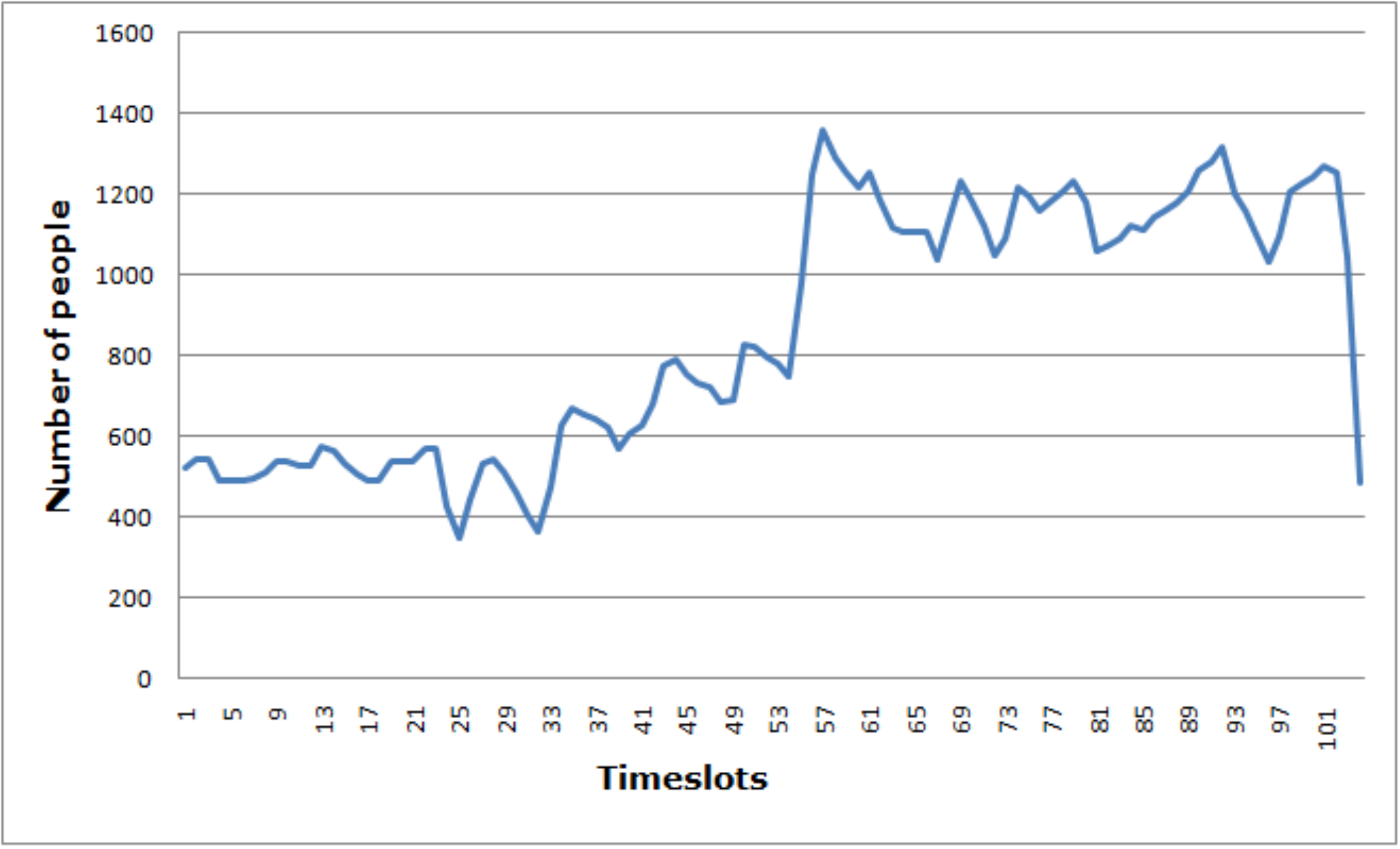}
\label{fig:member1}
}
\subfloat[2 and 3 groups]{
\includegraphics[scale=0.28]{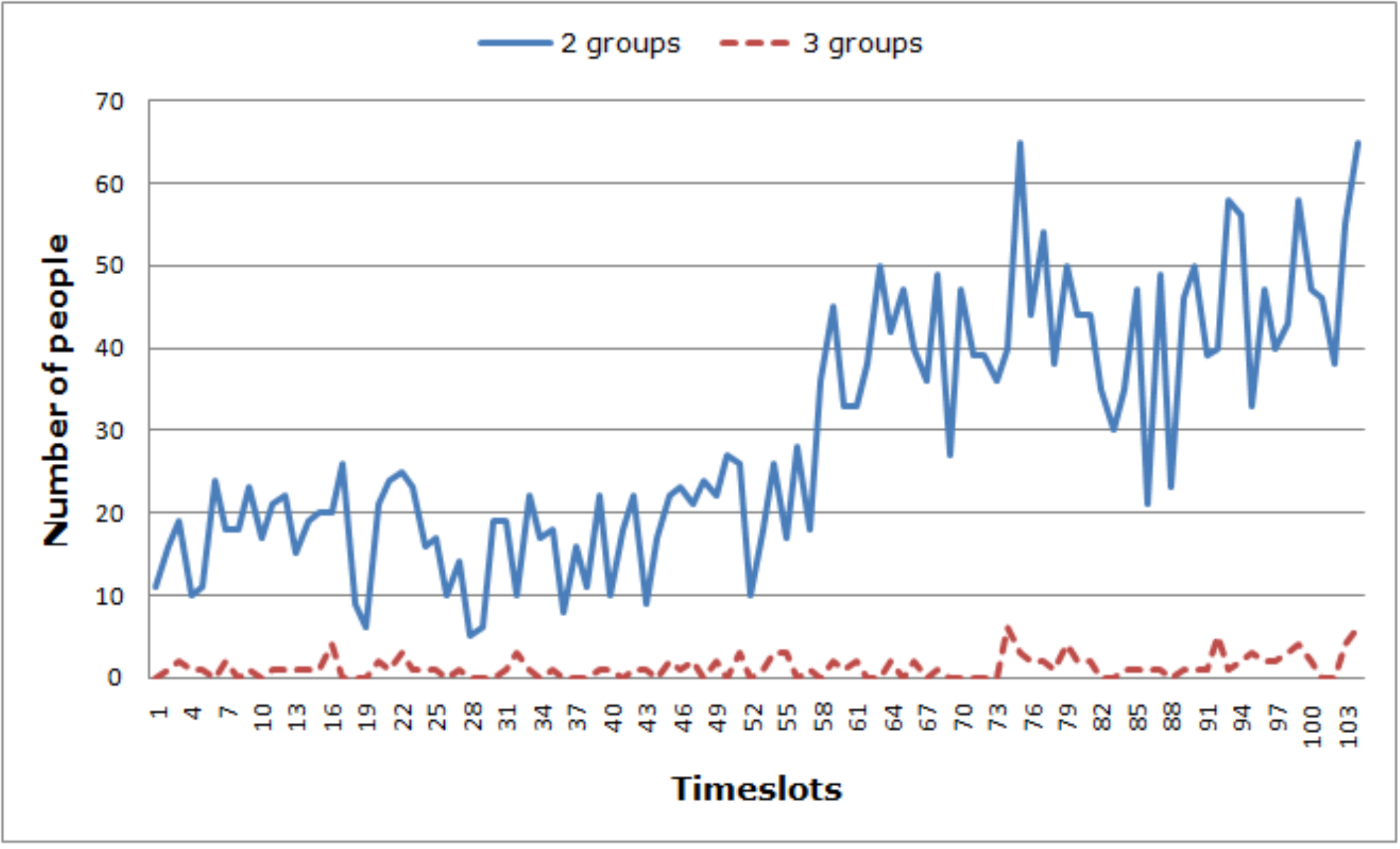}
\label{fig:member2_3}
}
\caption{Membership of people to groups for k=3 in comments model.}
\end{figure}

In figs. \ref{fig:member1} and \ref{fig:member2_3}, the numbers of users belonging to one, two or three stable groups in each interval for k=3 are specified. The figure presents mentioned belongings only in the comments model, but in the post model diagram is very similar. We can notice that these numbers increase, mostly because of the increase of the popularity of the portal and the significance of political events taking place.

In tab. \ref{tab:stableGroupsPostComments} there are presented the total numbers of stable groups with different sizes, calculated for k equal 3, 4 and 5, for models based on comments assigned to post author ($\psi_{pn}$) and previous comments authors ($\psi_{cn}$).
The most significant differences are obtained for low sizes of groups. Usually, models with comments give more groups, because of higher quantity of different links in these models.

\begin{table}\scriptsize
\caption{Comparison between posts and comments models. }
\subfloat[Stable group sizes]{
\label{tab:stableGroupsPostComments}
\begin{tabular}{rrrrrrr}
\cline{2-7}
\noalign{\smallskip}
& \multicolumn{2}{l}{k=3} & \multicolumn{2}{l}{k=4} & \multicolumn{2}{l}{k=5} \\
\hline
\noalign{\smallskip}
\multicolumn{1}{l}{Size} & post & \multicolumn{1}{l}{comments} & post & \multicolumn{1}{l}{comments} & post & \multicolumn{1}{l}{comments}\\
\noalign{\smallskip}
\hline
\noalign{\smallskip}
3 & 992 & 1350 & 0 & 0 & 0 & 0\\
4 & 81 & 147 & 1373 & 1358 & 0 & 0\\
5 & 25 & 32 & 235 & 210 & 966 & 1059\\
6 & 7 & 10 & 52 & 54 & 213 & 205\\
7 & 2 & 4 & 19 & 21 & 74 & 63\\
8 & 1 & 3 & 13 & 6 & 39 & 35\\
9 & 3 & 1 & 3 & 10 & 26 & 26\\
10 & 0 & 0 & 5 & 4 & 20 & 13\\
11-50 & 0 & 0 & 40 & 58 & 50 & 121\\
51-100 & 0 & 0 & 1 & 0 & 14 & 10\\
101-200 & 0 & 0 & 4 & 5 & 30 & 22\\
$>200$ & 104 & 104 & 98 & 99 & 57 & 69\\
\hline
\end{tabular}}
\quad
\subfloat[Mean values for stable groups]{
\label{tab:measuresGroupsPostComments}
\begin{tabular}{llllll}
\hline\noalign{\smallskip}
Measure & Model & k=3 & k=4 & k=5\\
\noalign{\smallskip}
\hline
\noalign{\smallskip}
\multirow{2}{*}{Stability} & \multirow{1}{*}{post} & 0.100 & 0.081 & 0.099 \\
& \multirow{1}{*}{comments} & 0.133 & 0.098 & 0.106 \\
\hline
\multirow{2}{*}{Density} & \multirow{1}{*}{post} & 0.459 & 0.489 & 0.511 \\
& \multirow{1}{*}{comments} & 0.598 & 0.631 & 0.657 \\
\hline
\multirow{2}{*}{Cohesion} & \multirow{1}{*}{post} & 73.7 & 36.5 & 29.8 \\
& \multirow{1}{*}{comments} & 157.9 & 46.0 & 41.9 \\
\hline
\end{tabular}}
\vspace{-0.1cm}
\end{table}

In tab. \ref{tab:measuresGroupsPostComments} one can see, that comments model gives us more stable, dense and cohesive groups what is confirmed by their mean values. The comment model gives more different connected pairs of bloggers both inside the group which influence increase of density and cohesion. 

\subsection{Comparison of sentiment models}

In the next analysis we focused our attention on comparing models with comments without ($\psi_{cn}$) and with sentiment (different versions of ($\psi_{cs}$) function, described in section \ref{BuildingOfSocialRelations}) for k=3.

\begin{figure}[ht]
\centering
\subfloat[Sentiment counting model.]{
\includegraphics[scale=0.28]{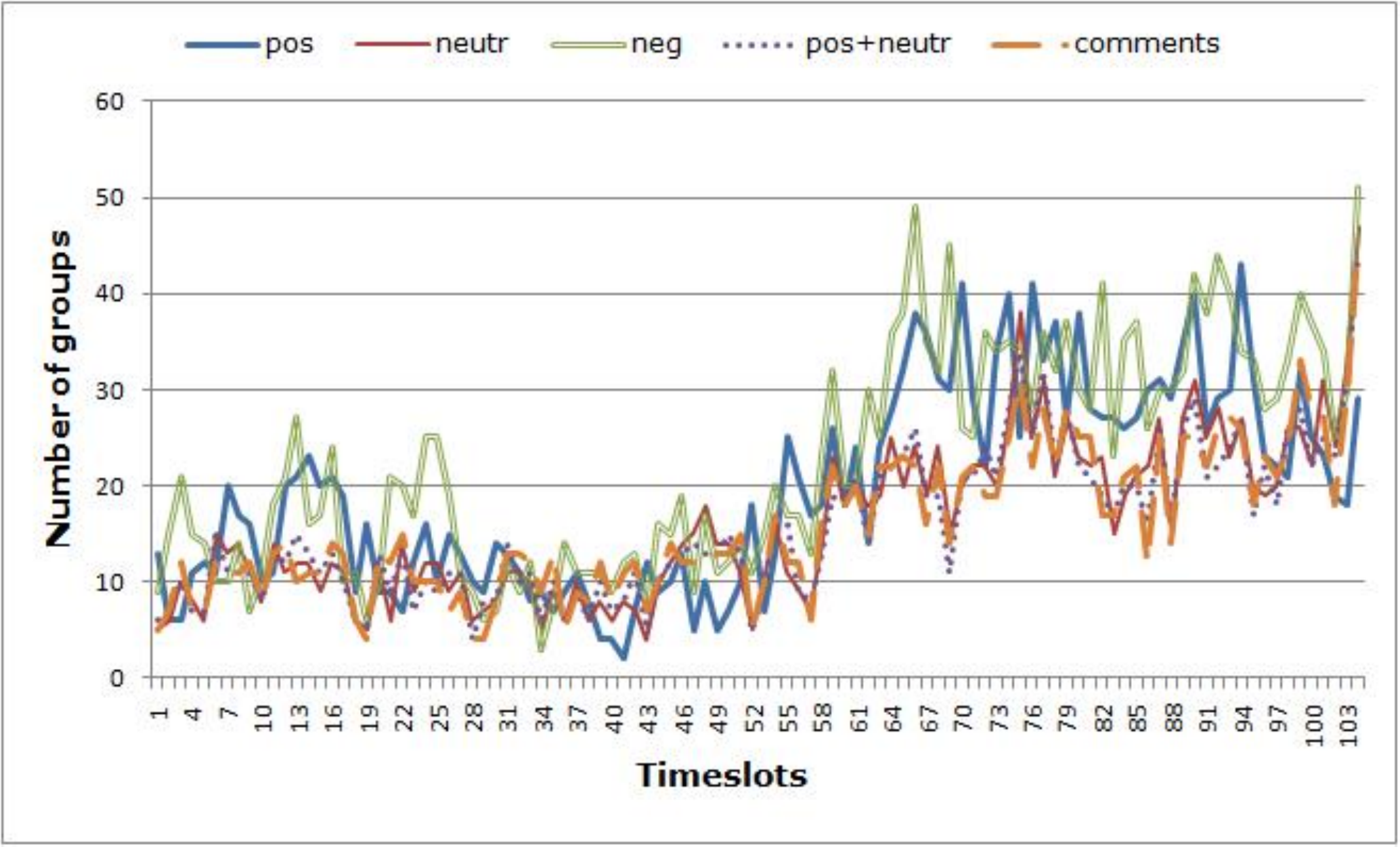}
\label{fig:grInSlotsSentCounting}
}
\subfloat[Sentiment mean model.]{
\includegraphics[scale=0.28]{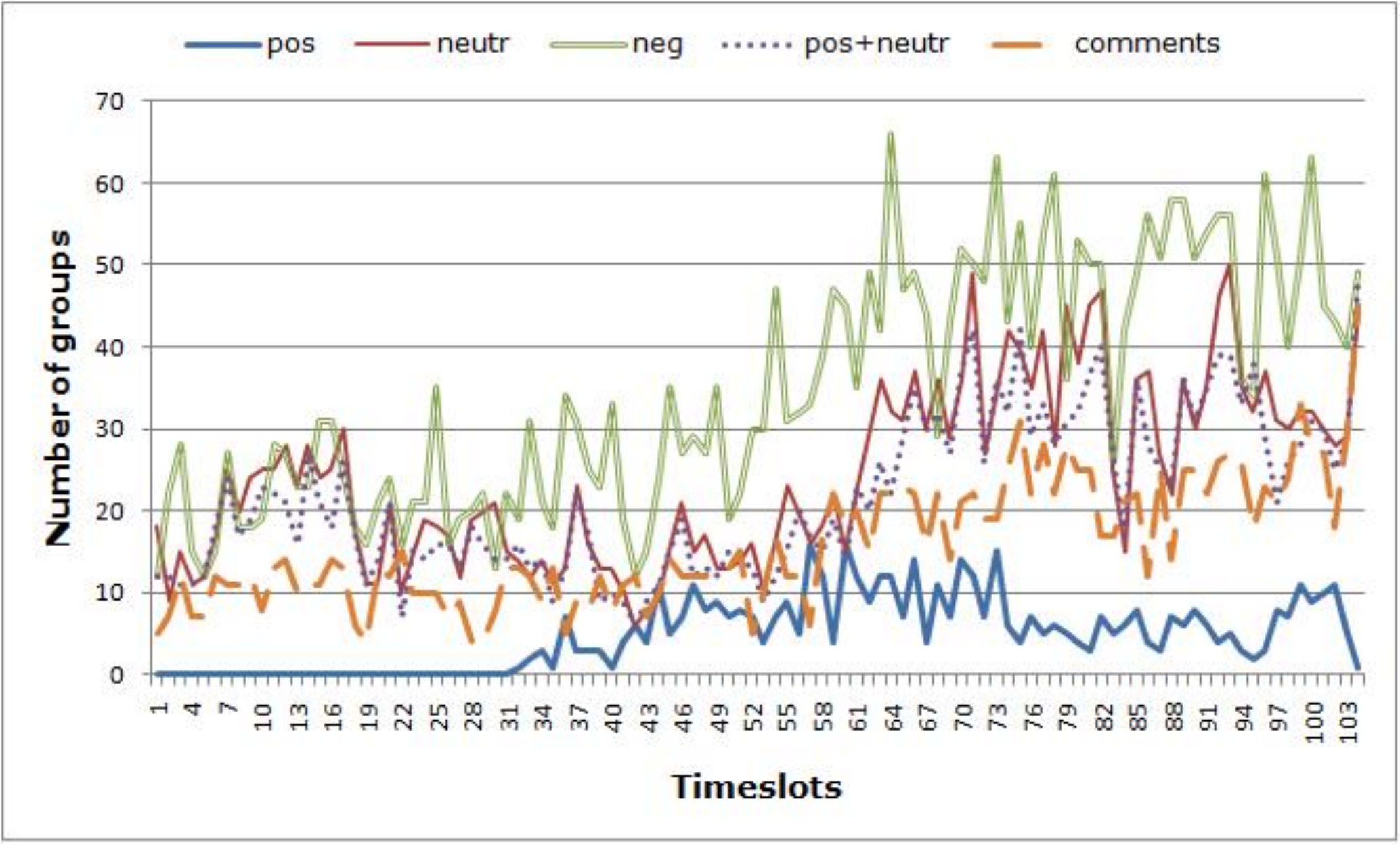}
\label{fig:grInSlotsSentMean}
}
\caption{Comparison of number of groups in slots in sentiments models for k=3.}
\end{figure}

In fig. \ref{fig:grInSlotsSentCounting} and \ref{fig:grInSlotsSentMean} the negative groups are dominating, but for groups in model with average sentiment (in fig. \ref{fig:grInSlotsSentMean} -- $\psi^a_{cs,n}$), stronger negative interactions are necessary to form them. One can notice, that such relations build well-shaped groups with strongly connected members.
Such behavior seems to be natural in the politic blogs, especially discussing controversial, emotion inspiring/arousing subjects. It is worth noting that negative relation between bloggers does not need to signify that the first blogger has a negative attitude regarding the second one, but that during the discussed subject they express negative emotions caused by another blogger or the general situation.

\begin{figure}[ht]
\centering
\subfloat[Sentiment counting model.]{
\includegraphics[scale=0.34]{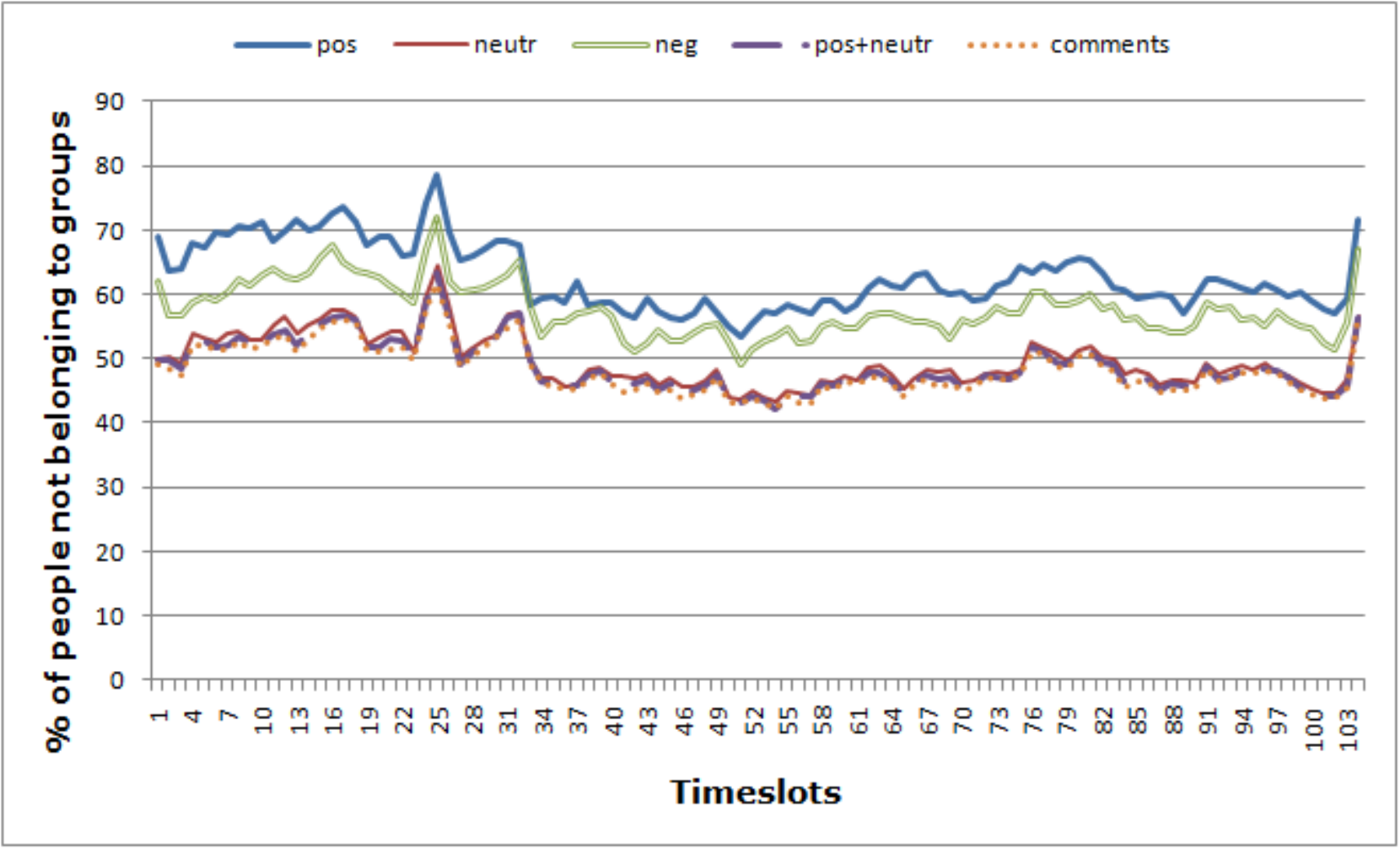}
\label{fig:notInGrSentCounting}
}
\subfloat[Sentiment mean model.]{
\includegraphics[scale=0.34]{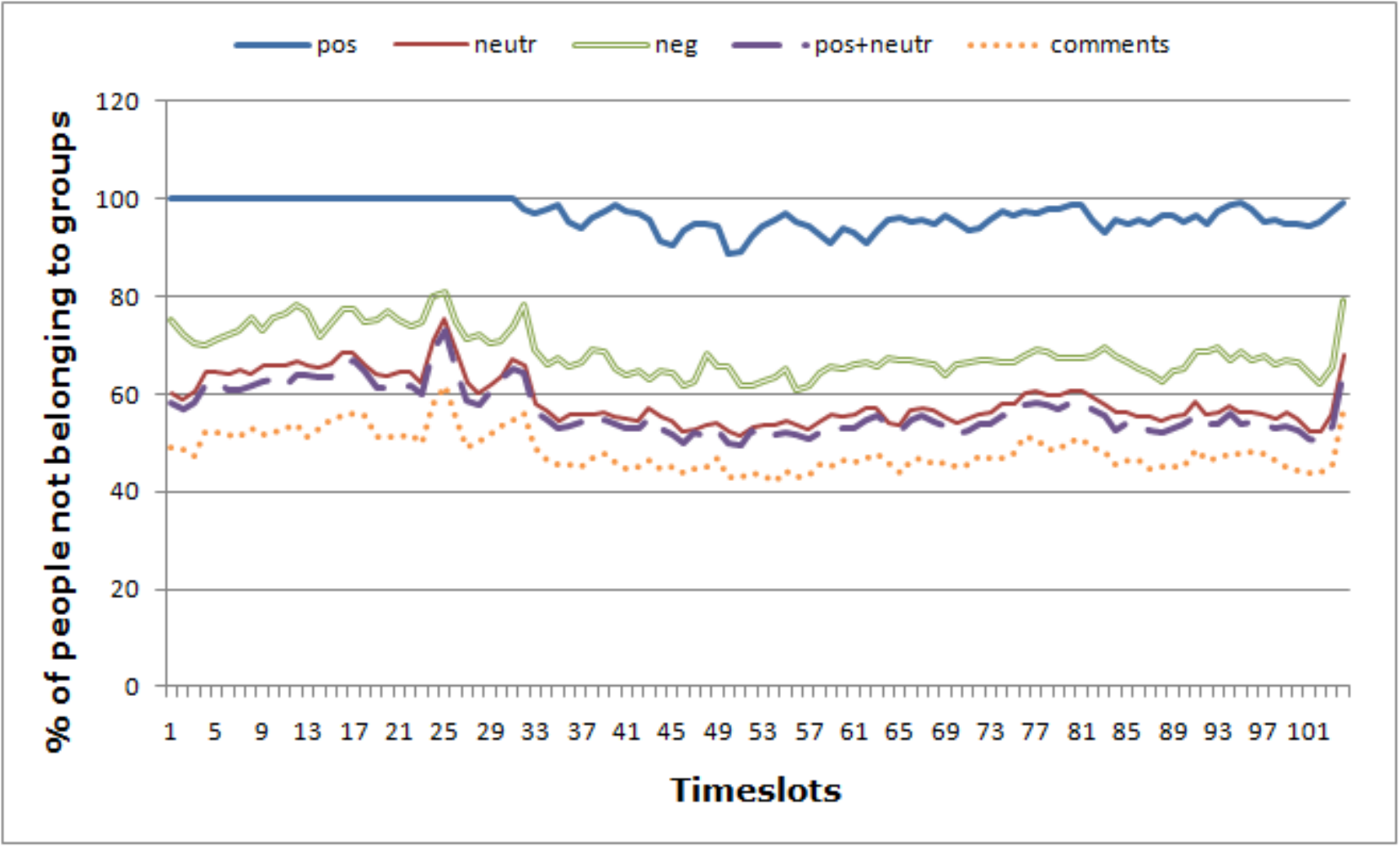}
\label{fig:notInGrSentMean}
}
\caption{Comparison of percent of users not belonging to any stable group in sentiment models for k=3.}
\vspace{-0.1cm}
\end{figure}

In fig. \ref{fig:notInGrSentCounting} and fig. \ref{fig:notInGrSentMean} one can see a significant difference between models when counting each kind of sentiment interactions separately and using the average value of the sentiment.  In mean comments model interactions with positive and negative sentiment canceling each other out and the obtained average is close to 0, for this reason there are significantly more persons belonging to the groups constructed for neutral average sentiment ($\psi^a_{cs,i}$).
It confirms the predictions that a model with an average value of sentiment identifies only radical sentiments in the case of positive and negative relations.

\begin{table}\scriptsize
\caption{Comparison of stable groups sizes between sentiment models for k=3. \label{tab:groupSizesSentiment}}
\subfloat[Sentiment counting model]{
\begin{tabular}{rrrrrr}
\hline\noalign{\smallskip}
Size & pos & neutr & neg & pos+neutr & comments\\
\noalign{\smallskip}
\hline
\noalign{\smallskip}
3 & 1606 & 1359 & 1855 & 1329 & 1350 \\
4 & 220 & 147 & 256 & 149 & 147 \\
5 & 52 & 33 & 74 & 36 & 32 \\
6 & 19 & 12 & 14 & 10 & 10 \\
7 & 16 & 2 & 11 & 4 & 4 \\
8 & 5 & 2 & 8 & 2 & 3\\
9 & 4 & 3 & 4 & 2 & 1 \\
10 & 2 & 1 & 3 & 0 & 0 \\
11-50 & 10 & 3 & 13 & 2 & 0 \\
51-100 & 0 & 0 & 0 & 0 & 0 \\
101-200 & 14 & 0 & 2 & 0 & 0 \\
$>200$ & 90 & 104 & 102 & 104 & 104 \\
\hline
\end{tabular}}
\quad
\subfloat[Sentiment mean model]{
\begin{tabular}{rrrrrr}
\hline\noalign{\smallskip}
Size & pos & neutr & neg & pos+neutr & comments\\
\noalign{\smallskip}
\hline
\noalign{\smallskip}
3 & 282 & 2071 & 2780 & 1873 & 1350 \\
4 & 87 & 238 & 458 & 201 & 147 \\
5 & 25 & 58 & 139 & 57 & 32 \\
6 & 21 & 23 & 62 & 16 & 10 \\
7 & 15 & 10 & 22 & 9 & 4 \\
8 & 6 & 3 & 18 & 0 & 3\\
9 & 6 & 3 & 7 & 1 & 1 \\
10 & 6 & 2 & 12 & 3 & 0 \\
11-50 & 48 & 12 & 20 & 6 & 0 \\
51-100 & 3 & 0 & 2 & 0 & 0 \\
101-200 & 0 & 1 & 28 & 1 & 0 \\
$>200$ & 0 & 103 & 72 & 103 & 104 \\
\hline
\end{tabular}}
\vspace{-0.8cm}
\end{table}

\begin{table}\scriptsize
\caption{Comparison of stable groups parameters (mean values for all stable groups in time slots) between sentiment models for k=3. \label{tab:MeasuresSentimentModels}}
\subfloat[Sentiment counting model]{
\begin{tabular}{rrrrrr}
\hline\noalign{\smallskip}
Model & Stability & Density & Cohesion\\
\noalign{\smallskip}
\hline
\noalign{\smallskip}
pos & 0.114 & 0.538 & 89.8 \\
neutr & 0.130 & 0.59 & 157.3 \\
neg & 0.117 & 0.557 & 135.4 \\
pos+neutr & 0.135 & 0.593 & 157.4 \\
comments & 0.133 & 0.598 & 157.9 \\
\hline
\end{tabular}}
\subfloat[Sentiment mean model]{
\begin{tabular}{rrrrrr}
\hline\noalign{\smallskip}
Model & Stability & Density & Cohesion\\
\noalign{\smallskip}
\hline
\noalign{\smallskip}
pos & 0.229 & 0.448 & 34.8 \\
neutr & 0.087 & 0.545 & 104.8 \\
neg & 0.087 & 0.526 & 61.4 \\
pos+neutr & 0.097 & 0.554 & 116.6 \\
comments & 0.133 & 0.598 & 157.9 \\
\hline
\end{tabular}}
\end{table}

Analyzing the total number of groups with different sizes depending on model and character of polarization (tab. \ref{tab:groupSizesSentiment}), one can notice that counting separately the groups in each model, the sentiment counting models give much more positives groups then sentiment mean models, but significantly less for negative and neutral groups. 

In the sentiment mean model the most stable groups were obtained for positive sentiment (tab. \ref{tab:MeasuresSentimentModels}), it may be caused by the fact, that the number of these groups is low (as can be seen in tab. \ref{tab:groupSizesSentiment}). The method of the identification of relations used in this model gave only groups exchanging very positive content, such specific groups are characterized by a high stability of memberships.     
For remaining models, measures of groups for sentiment mean models are lower or much lower than for the sentiment counting models, so they identify groups less dense, less stable and less separated from the environment. In sentiment mean model there is a lot less connections between nodes 
than in sentiment counting model, so it may explain smaller values of density. 

\section{Conclusion}

The paper introduces a set of developed models describing social networks, taking into consideration different kinds of interactions and sentiment polarization. Models were applied to the analysis of stable groups, identified in the selected blog portal. The introduced set of models can help in systematization of the problem domain and allow us to identify research directions and relations between them. 

Several experiments were conducted which delivered new, detailed information about a character and behavior of groups of users on the portal.
The method of identification of stable groups in blogosphere was improved which allowed us to obtain more stable, dense and cohesive groups. In new model (comments model) lower number of users did not belong to any group.
Introduction of the sentiment as an interaction attribute allowed to observe different characteristic behaviors of groups with different polarization.
Positive sentiment groups are formed around not controversial topics while negative sentiment groups are associated with controversial matters and possibly quarrels.

The presented solutions will be applied to analyze other blog portals and different kinds of social media, for example microblogs.
The next works will embrace: improving the quality of the sentiment analysis, key bloggers identification and analysis of their memberships in given groups.
 We are going to integrate presented sentiment models with our research on group dynamics and prediction of group evolution, as well as the identification of the most significant, strongly linked members of the group, constituting group cores.
Another direction is to associate models based on sentiment with extended description of groups which considers the most popular discussed subjects identified by analysis of tags or post and comment content.

{\bf Acknowledgments.}
The authors thank P. Macio{\l}ek who provided and allowed the use of the algorithm and tools for analysis of sentiment of texts in Polish language.

\bibliographystyle{splncs03}
\bibliography{biblio,blogs,groups,snajkaz}

\begin{thebibliography}{10}
\providecommand{\url}[1]{\texttt{#1}}
\providecommand{\urlprefix}{URL }

\bibitem{AgarwalLiu:09}
Agarwal, N., Liu, H.: Modeling and Data Mining in Blogosphere. Moegan \&
  Claypool Publishers (2009)

\bibitem{Akritidis:2009}
Akritidis, L., Katsaros, D., Bozanis, P.: Identifying influential bloggers:
  Time does matter. In: Procs. of the 2009 IEEE/WIC/ACM Int. Joint Conf. on Web
  Intelligence and Intelligent Agent Technology - Vol. 01. pp. 76--83.
  WI-IAT'09, IEEE Computer Society, Washington, DC, USA (2009)

\bibitem{Bermingham2009}
Bermingham, A., Conway, M., McInerney, L., O'Hare, N., Smeaton, A.F.: Combining
  social network analysis and sentiment analysis to explore the potential for
  online radicalisation. In: Proc. of the 2009 Int. Conf. on Adv. in Social
  Network Analysis and Mining. pp. 231--236. IEEE Comp. Soc., Washington, DC,
  USA (2009)

\bibitem{Chi:2007}
Chi, Y., Zhu, S., Song, X., Tatemura, J., Tseng, B.L.: Structural and temporal
  analysis of the blogosphere through community factorization. In: Proc. of the
  13th ACM SIGKDD international conference on Knowledge discovery and data
  mining. pp. 163--172. KDD '07, ACM, New York, NY, USA (2007)

\bibitem{Fortunato2010}
Fortunato, S.: Community detection in graphs. In: Phys. Rep., chap. 486 (2010)

\bibitem{GliwaSZBKK12}
Gliwa, B., Saganowski, S., Zygmunt, A., Br{\'o}dka, P., Kazienko, P.,
  Ko{\'z}lak, J.: Identification of group changes in blogosphere. In: IEEE/ACM
  International Conference on Advances in Social Networks Analysis and Mining,
  ASONAM 2012 Istanbul, Turkey, 26-29 August 2012. IEEE Computer Society (2012)

\bibitem{Gryc2010}
Gryc, W., Moilanen, K.: Leveraging textual sentiment analysis with social
  network modelling: Sentiment analysis of political blogs in the 2008 u.s.
  presidential election. In: Procs. of the "From Text to Political Positions"
  Workshop (T2PP 2010). Vrije Universiteit, Amsterdam (2010)

\bibitem{KozlakZ11}
Ko{\'z}lak, J., Zygmunt, A.: Agent-based modelling of social organisations. In:
  International Conference on Complex, Intelligent and Software Intensive
  Systems, CISIS 2011, June 30 - July 2, 2011, Korean Bible University, Seoul,
  Korea. pp. 467--472. IEEE Computer Society (2011)

\bibitem{Kraus2008}
Krauss, J., Nann, S., Simon, D., Fischbach, K., Gloor, P.: Predicting movie
  success and academy awards through sentiment and social network analysis. In:
  ECIS 2008 Proceedings (2008)

\bibitem{Nguyen2010}
Nguyen, T., Phung, D.Q., Adams, B., Tran, T., Venkatesh, S.: Hyper-community
  detection in the blogosphere. In: Proceedings of second ACM SIGMM workshop on
  Social media. pp. 21--26. WSM '10, ACM, New York, NY, USA (2010)

\bibitem{Ning:2010}
Ning, H., Xu, W., Chi, Y., Gong, Y., Huang, T.S.: Incremental spectral
  clustering by efficiently updating the eigen-system. Pattern Recogn.  43(1),
  113--127 (2010)

\bibitem{palla2008}
Palla, G., Ábel, D., Farkas, I.J., Pollner, P., Derényi, I., Vicsek:, T.:
  k-clique percolation and clustering. In: Bollobás, B., Kozma, R., Miklós, D.
  (eds.) Handbook of Large-scale Random Networks. Springer (2009)

\bibitem{palla2005}
Palla, G., Derenyi, I., Farkas, I., Vicsek, T.: Uncovering the overlapping
  community structure of complex networks in nature and society. Nature  435,
  814--818 (2005)

\bibitem{Pang2008}
Pang, B., Lee, L.: Opinion mining and sentiment analysis. Foundations and
  Trends in Information Retrieval  2(1-2) (2008)

\bibitem{Shamma2010}
Shamma, D.A., Kennedy, L., Churchill, E.F.: Statler: Summarizing media through
  short-messaging services. In: CSW 2010. ACM, USA (2010)

\bibitem{Tang2010}
Tang, L., Liu, H.: Graph mining applications to social network analysis. In:
  Aggarwal, C., Wang, X. (eds.) Managing and Mining Graph Data. Springer (2010)

\bibitem{Tromp2011}
Tromp, E., Pechenizkiy, M.: Senticorr: Multilingual sentiment analysis of
  personal correspondence. In: Proc. of ICDM 2011 Workshops. IEEE Press (2011)

\bibitem{Xu2011}
Xu, K., Li, J., Liao, S.S.: Sentiment community detection in social networks.
  In: Procs. of the 2011 iConference. pp. 804--805. iConf. '11, ACM, NY, USA
  (2011)

\bibitem{ZygmuntBKK11}
Zygmunt, A., Br{\'o}dka, P., Kazienko, P., Ko{\'z}lak, J.: Different approaches
  to groups and key person identification in blogosphere. In: International
  Conference on Advances in Social Networks Analysis and Mining, ASONAM 2011,
  Kaohsiung, Taiwan, 25-27 July 2011. pp. 593--598. IEEE Computer Society
  (2011)

\bibitem{ZygmuntBKK12}
Zygmunt, A., Br{\'o}dka, P., Kazienko, P., Ko{\'z}lak, J.: Key person analysis
  in social communities within the blogosphere. J. UCS  18(4),  577--597 (2012)

\end{thebibliography}

\end{document}